\documentclass[aps,pre,twocolumn,floats,amsmath,amssymb,showpacs]{revtex4}

\usepackage{graphicx}
\usepackage{dcolumn}
\usepackage{bm}
\usepackage{epsfig}
\usepackage{float}
\usepackage{gensymb}
\usepackage{subfigure}
\usepackage{latexsym}
\usepackage{ragged2e}
\usepackage{textcomp}
\usepackage{gensymb}
\usepackage{etoolbox}
\usepackage{color}

\begin{document}
\newcommand{\hide}[1]{}
\newcommand{\tbox}[1]{\mbox{\tiny #1}}
\newcommand{\half}{\mbox{\small $\frac{1}{2}$}}
\newcommand{\sinc}{\mbox{sinc}}
\newcommand{\const}{\mbox{const}}
\newcommand{\trc}{\mbox{trace}}
\newcommand{\intt}{\int\!\!\!\!\int }
\newcommand{\ointt}{\int\!\!\!\!\int\!\!\!\!\!\circ\ }
\newcommand{\eexp}{\mbox{e}^}
\newcommand{\bra}{\left\langle}
\newcommand{\ket}{\right\rangle}
\newcommand{\EPS} {\mbox{\LARGE $\epsilon$}}
\newcommand{\ar}{\mathsf r}
\newcommand{\im}{\mbox{Im}}
\newcommand{\re}{\mbox{Re}}
\newcommand{\bmsf}[1]{\bm{\mathsf{#1}}}
\newcommand{\mpg}[2][1.0\hsize]{\begin{minipage}[b]{#1}{#2}\end{minipage}}

\title{Multilayer directed random networks: Scaling of spectral properties}

\author{G. Tapia-Labra,$^1$ M. Hern\'andez-S\'anchez,$^1$ and J. A. M\'endez-Berm\'udez$^{1,2}$}
\address{Instituto de F\'isica, Benem\'erita Universidad Aut\'onoma de Puebla, Puebla 72570, Mexico \\
Escuela de F\'isica, Facultad de Ciencias, Universidad Nacional Aut\'onoma de Honduras, Honduras}

\date{\today}

\begin{abstract}
Motivated by the wide presence of multilayer networks in both natural and human-made systems,
within a random matrix theory (RMT) approach, in this study we compute eigenfunction and spectral  
properties of multilayer directed random networks (MDRNs) in two setups composed by $M$ layers 
of size $N$: A line and a complete graph (node-aligned multiplex network).
First, we numerically demonstrate that the normalized localization length $\beta$ of the eigenfunctions 
of MDRNs follows a simple scaling law given by $\beta=x^*/(1+x^*)$, with 
$x^*\propto (b_{\mbox{\tiny eff}}^2/L)^\delta$, $\delta\sim 1$ and $b_{\mbox{\tiny eff}}$ being 
the effective bandwidth of the adjacency matrix of the network of size $L=M\times N$. 
Here, $b_{\mbox{\tiny eff}}$ incorporates both intra-- and inter--layer edges.
Then, we show that other eigenfunction and spectral RMT measures 
(the inverse participation ratio of eigenfunctions, the ratio between nearest- and next-to-nearest-
neighbor eigenvalue distances, and the ratio between consecutive singular-value spacings) of MDRNs 
also scale with $x^*$.
\end{abstract}


\maketitle

\section{Introduction}

A network is a collection of nodes interconnected by links. In a single-layer network, nodes are connected through only one type of link. However, in many real systems different types of interactions and relationships between different agents (nodes) may exist. These relationships are better represented by a multilayer or multiplex network, see e.g.~\cite{BoccalettiPR2014,Kivela2014,DeDomenico2013,Cozzo2015}.
A multilayer network consists of multiple layers, each layer representing a different type of relationship between nodes. For example, in a social network, one layer may represent friendship relationships, while another may represent professional relationships. In a transportation network, one layer may represent road connections between cities, while another may represent railway connections between cities. Then, nodes in one layer can be connected to nodes on other layers.

Using multilayer and multiplex network analysis, one could better understand the underlying structure of complex systems, identify influential nodes and links, and predict how the corresponding system will be affected by changes in its structure or behavior. This type of analysis is important for understanding how networked real systems work and how they can be optimized or improved.

On the other hand, a directed network is a type of graph in which links have a direction associated with them~\cite{BG00,BG18}. Each link has a source node and a target node, and the directionality of a link indicates the flow of influence or information from its source node to its target node.
In directed networks, the links are not symmetrical, meaning that if A is directly connected to B, it does not automatically follow that B is directly connected to A. This asymmetry allows for a better understanding of the flow of influence in complex systems, such as social networks and biological networks.
Evidently, multilayer networks can also incorporate the property of directionality. 
However, even when most real systems that can be represented by multilayer networks are intrinsically directed, directed multilayer network models have not been widely studied yet; for few very recent exceptions, see~\cite{RJ23,SCGY24,Q24,WLW24,LLL23,WZ22,GA24}.
Therefore, in order to contribute to fill this gap, in this work we perform a detailed numerical study of eigenfunction and spectral  
properties of multilayer directed random networks within a random matrix theory approach.

Indeed, Random Matrix Theory (RMT) has numerous applications in many different fields, from condensed matter 
physics to financial markets~\cite{RMT}. 
In the case of complex networks, the use of RMT techniques might reveal universal properties.
For instance, it has been shown that
(i) the eigenvalue statistics of the adjacency matrices of several models of random 
networks (in the dense limit) follows the statistics of the standard RMT ensembles 
(see e.g.~\cite{Bandyopadhyay,MAM15,PRRCM20,PM23,MA24,RJ23}) while
(ii) the corresponding eigenfunction properties show 
a localization-to-delocalization crossover (see e.g.~\cite{MAM15,PRRCM20,PM23,RJ23}), 
as a function of the average degree, proper of RMT parametric ensembles.
Indeed, some models of directed random networks and graphs have already been studied 
by the use of RMT models and techniques; as examples we can mention directed 
Erd\"os-R\'enyi networks~\cite{PRRCM20,MA24,MMS24}, 
directed random geometric graphs~\cite{PM23,MA24}, and 
directed multilayer networks~\cite{RJ23}.

Admittedly, we can identify two papers where eigenfunction and spectral  
properties of multilayer random networks have been reported by the use of a RMT approach:
one where the multilayer random networks are non-directed~\cite{MFRM17} and the other where the 
directed multilayer networks consist of two layers only~\cite{RJ23}.
In this respect, the present study is more general since it introduces a RMT null model for the adjacency 
matrices of of directed multilayer random networks where the number of layers enters as a model
parameter.

\section{Model definitions and RMT measures}

\begin{figure}[t]
\centerline{\includegraphics[width=\columnwidth]{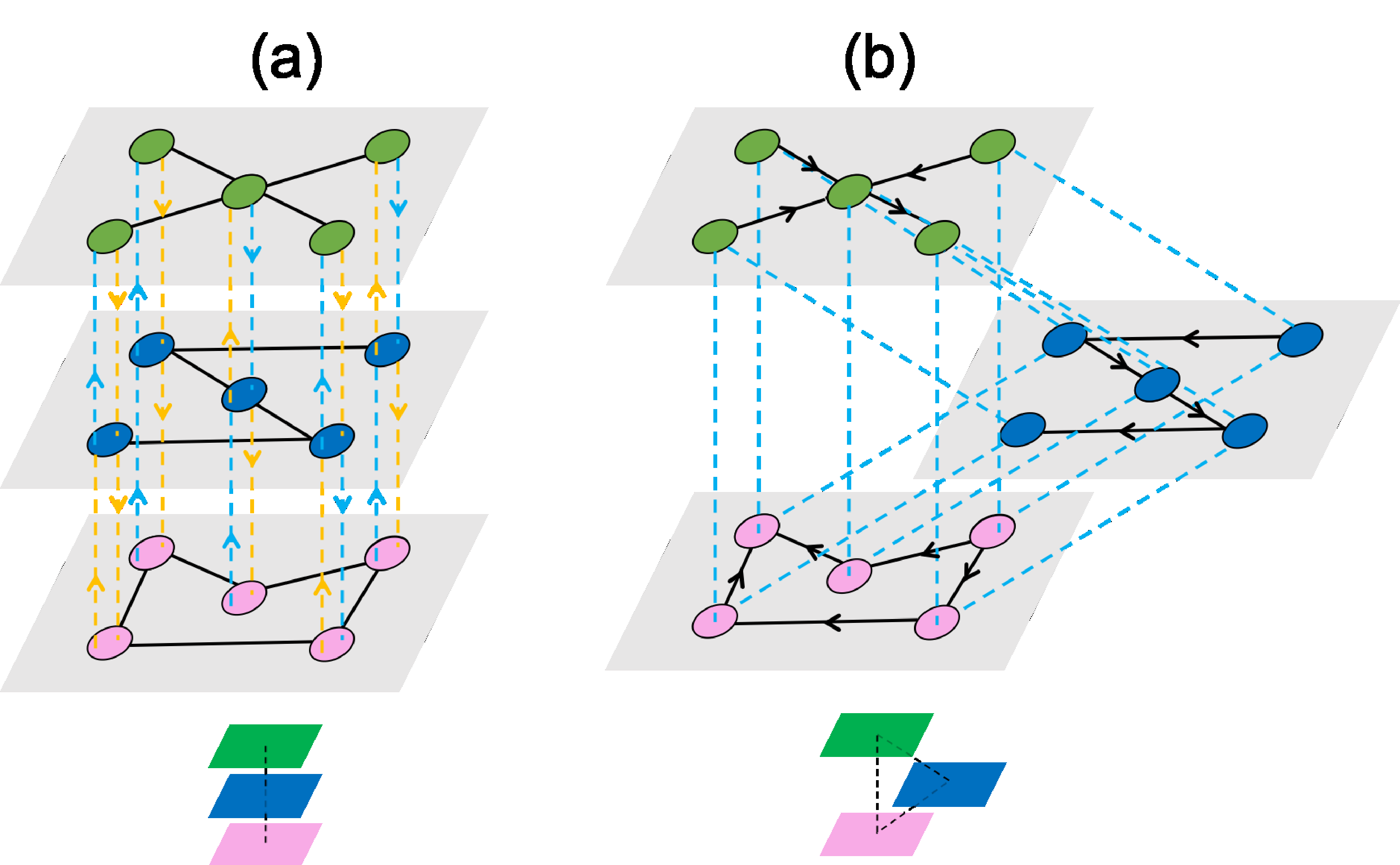}}
\caption{Illustration of the two multilayer directed random networks studied here. 
The network of layers are (a) a line and (b) a complete network. 
Here, each directed network is composed by $M=3$ layers having $N=5$ nodes each.
In (a) the interlayer edges are undirected, while the edges between layers are directed.
In contrast, in (b) the interlayer edges are directed, while the edges between layers are undirected.}
\label{Fig00}
\end{figure}

\subsection{Models of multilayer directed random networks}
 
The multilayer directed network we consider here is formed by $M$ random networks (layers) with
corresponding adjacency matrices $A^{(m)}$ having $N_m$ nodes each. 
The adjacency matrix of the whole network is expressed by 
${\bf A} = \bigoplus_{m=1}^M A^{(m)} + \omega \bf{C}$, where $\bigoplus$ represents the Kronecker
product, $\omega$ is a parameter that defines the strength of the interlayer edges and $\bf{C}$  is the 
interlayer coupling matrix, whose elements represent the relations between nodes in different layers. 
Examples of directed multilayers are shown in Fig.~\ref{Fig00}. 
Notice that for $\omega \ll 1$ the layers can be considered as uncoupled, while for $\omega \gg 1$ the 
topology of the network of layers dominates the spectral properties~\cite{Garcia2014,Cozzo2016}. In 
this way, $\omega = 1$ represents a suitable intermediary case (multilayer phase). 
In what follows, we focus on the case $\omega =1$, moreover in Appendix~\ref{app} we explore the 
situation of $\omega \ne 1$. 

In this study we define two ensembles of directed multilayer random networks as adjacency matrices. 

As the first model 
we consider a network of layers on a line, see Fig.~\ref{Fig00}(a), whose adjacency matrix 
${\bf A}$ has the form 
\begin{equation}
{\bf A} =
\left(
\begin{array}{ccccc}
A^{(1)} & C^{(1,2)} & 0 & \cdots & 0 \\
C^{(2,1)} & A^{(2)} & C^{(2,3)} & \ & 0 \\
0 & C^{(3,2)} & A^{(3)} & \ & 0 \\
\vdots & \ & \ & \ddots & C^{(M-1,M)} \\
0 & 0 & 0 & C^{(M,M-1)} & A^{(M)}
\end{array}
\right) \ ,
\label{eq:A_line}
\end{equation}
where $C^{(m,m')}$ are real and, in general, rectangular 
matrices of size $N_m \times N_{m'}$ and 0 represents null matrices. Furthermore, we consider a 
special class of matrices $A^{(m)}$ and $C^{(m,m')}$ which are characterized by the sparsities 
$\alpha_A$ and $\alpha_C$, respectively. In other words, since with a probability $\alpha_*$ their 
elements can be removed, these matrices represent Erd\"os-R\'enyi--type random networks. 
For this setup we consider the interlayer edges as undirected, while the edges between layers 
are directed; that is, $\left(A^{(m,m')}\right)_{i,j}=\left(A^{(m,m')}\right)^{\mbox{\tiny T}}_{j,i}$
and $\left(C^{(m,m')}\right)_{i,j} \ne \left(C^{(m,m')}\right)^{\mbox{\tiny T}}_{j,i}$.
The corresponding undirected case has been studied in Ref.~\cite{MGRM17}. Notice 
that when the $N_m$ are all the same $N_m=\mbox{constant}\equiv N$, which is the case we 
explore  here. Under this assumption, the adjacency matrix ${\bf A}$ has the structure of a 
block-banded matrix of 
size $L=M\times N$. In addition, we consider this model as a model of weighted networks; i.e., 
the non-vanishing elements ${\bf A}_{i,j}$ are independent Gaussian variables with 
zero mean and variance one. 
We justify the addition of self-loops and random weights to edges by recognizing
that in real-world networks the nodes and the interactions between them are in general
non-equivalent. According to this definition
diagonal random matrices are obtained for $\alpha_A=\alpha_C=0$, known as the Poisson 
ensemble (PE)~\cite{Metha} in RMT. For simplicity, and without loss 
of generality, in this work we consider the case where $\alpha\equiv\alpha_A=\alpha_C$.
As an example, this network model can be applied to transportation networks, where the
interlayer edges represent connections between two different means of transport.  
An obvious constraint of this setup is that no layer can be connected to more than two layers. 

As the second model we consider the node-aligned multiplex
case, whose network of layers is a complete graph, see Fig.~\ref{Fig00}(b).
In this setup the $M$ layers have the same number of nodes $N_m=\mbox{constant}\equiv N$
while the coupling matrices are restricted to identity matrices, $C^{(m,m')}=I$ of size $N\times N$. 
Then, the adjacency matrix of a node-aligned multiplex is given as
\begin{equation}
{\bf A} =
\left(
\begin{array}{ccccc}
A^{(1)} & I & I & \cdots & I \\
I & A^{(2)} & I & \ & I \\
I & I & A^{(3)} & \ & I \\
\vdots & \ & \ & \ddots & I \\
I & I & I & I & A^{(M)}
\end{array}
\right) \ .
\label{eq:A_mux}
\end{equation}
Similarly to the multilayer setup of Eq.~(\ref{eq:A_line}), this configuration is characterized by 
the sparsity $\alpha$ (i.e., each layer is an Erd\"os-R\'enyi--type random network)
which we choose to be constant for all the $M$ matrices $A^{(m)}$ of size $N\times N$ 
composing the adjacency matrix ${\bf A}$ of size $L=M\times N$. 
Also, the non-vanishing elements of the matrices $A^{(m)}$ are chosen as independent Gaussian 
variables with zero mean and variance one. 
For this setup we consider the inter-layer edges as directed, 
$\left(A^{(m,m')}\right)_{i,j} \ne \left(A^{(m,m')}\right)^{\mbox{\tiny T}}_{j,i}$, while the edges between 
layers are undirected. 
This is a quite natural setup since the coupling matrices were already defined as identity matrices.
The corresponding undirected case has been studied in Ref.~\cite{MGRM17}.
A realistic example of this configuration is an online 
social systems, where each layer represents a different online network (e.g., Facebook, X, 
Google+, etc).

\subsection{Random Matrix Theory measures}

We use standard RMT measures to characterize the eigenfunction and 
spectral properties of the non-Hermitian adjacency matrices ${\bf A}$ belonging to the two 
directed multilayer setups described above.

Regarding eigenfunction properties, given the normalized eigenfunctions $\Psi^i$ 
(i.e.~$\sum_{m=1}^L\vert\Psi_m^i\vert^2 = 1$) of ${\bf A}$, we compute 
the Shannon entropies~\cite{S48}
\begin{equation}
S_i = \sum_{m=1}^L\vert\Psi_m^i\vert^2 \ln\vert\Psi_m^i\vert^2
\label{shannon}
\end{equation}
and the inverse participation ratios~\cite{OH07}
\begin{equation}
\mbox{IPR}_i = \left[\sum_{m=1}^L\vert\Psi_m^i\vert^4\right]^{-1}.
\label{IPR}
\end{equation}
Both $S$ and $\mbox{IPR}$ measure the extension of eigenfunctions on a given basis. 

Regarding spectral properties, given the complex spectrum $\{ \lambda_i \}$ ($i=1\ldots L$) of the 
non-Hermitian adjacency matrix ${\bf A}$, we compute the ratio $r_{\mathbb{C}}$ between nearest- 
and next-to-nearest-neighbor eigenvalue distances, with the $i$--th ratio defined as~\cite{SRP20}
\begin{equation}
r_{\mathbb{C}}^i = \frac{\vert \lambda_i^{\mathrm{nn}}-\lambda_i\vert}{\vert \lambda_i^{\mathrm{nnn}}-\lambda_i\vert} ;
\label{rC}
\end{equation}
where $\lambda_i^{\mathrm{nn}}$ and $\lambda_i^{\mathrm{nnn}}$ are, respectively, the nearest and the 
next-to-nearest neighbors of $\lambda_i$ in $\mathbb{C}$.

Recently, the singular-value statistics (SVS) has been presented as a RMT tool able to properly 
characterize non-Hermitian RM ensembles~\cite{KXOS23} as well as to identify
the delocalization transition in non-Hermitian many-body systems~\cite{RBSC24} and models of 
directed networks~\cite{MA24}.
So, we also use SVS here to characterize 
spectral properties of ${\bf A}$ as follows: Given the ordered square roots of the real eigenvalues of 
the Hermitian matrix ${\bf A}{\bf A}^\dagger$, $s_1>s_2>\cdots >s_L$ (which are the singular 
values of ${\bf A}$), we compute the ratio $r_{\tbox{SV}}$ between consecutive singular-value spacings, 
where the $i$--th ratio is given by~\cite{KXOS23}
\begin{equation}
r_{\tbox{SV}}^i = \frac{\mathrm{min}(s_{i+1}-s_i,s_{i}-s_{i-1})}{\mathrm{max}(s_{i+1}-s_i,s_i-s_{i-1})} .
\label{rSV}
\end{equation}
Above, as usual, ${\bf A}^\dagger$ is the conjugate transpose of ${\bf A}$.
Moreover, since for real matrices, as the ones we consider here, the conjugate transpose is just the 
transpose ${\bf A}^\dagger={\bf A}^{\tbox{T}}$,
then, in what follows, the SVS concerns the spectra of ${\bf A}{\bf A}^{\tbox{T}}$.

In the following we use exact numerical diagonalization to obtain the right eigenfunctions 
$\Psi^i$ ($i=1\ldots L$), the complex eigenvalues $\lambda_i$, and the singular values $s_i$  
of large ensembles of non-Hermitian adjacency matrices ${\bf A}$ characterized by the parameter set ($M,N,\alpha$). 
For each of the averages reported below we used at least $5\times 10^5$ data values.

\section{Scaling analysis of multilayer directed random networks}

Particularly, Shannon entropies allows to compute the entropic eigenfunction localization 
length, see e.g.~\cite{I90},
\begin{equation}
\label{lH}
\ell_L = L \exp\left[ -\left( S_{\tbox{RGE}} - \bra S \ket \right)\right] ,
\end{equation}
where $S_{\tbox{RGE}}\approx\ln(L/1.56)$~\cite{PRRCM20} 
is the average entropy of the eigenfunctions of the real Ginibre ensemble (RGE)~\cite{G65}. 
We use $S_{\tbox{RGE}}$ as the reference entropy because, as well as the
adjacency matrices of our multilayer directed random networks, the RGE consists of
non-Hermitian real random matrices.
With this definition for $\ell_L$, when $\alpha=0$, ${\bf A}$ becomes a diagonal real random 
matrix (that is, a member of the Poisson Ensemble (PE)~\cite{Metha}) and the corresponding
eigenfunctions have only one non-vanishing component 
with magnitude equal to one; so $\bra S \ket=0$ and $\ell_L\sim 1$. On the contrary, 
when $\alpha=1$ and $M=2$ we recover the RGE and $\bra S \ket=S_{\tbox{RGE}}$; 
so, the {\it fully chaotic} eigenfunctions extend over the $L$ available basis states
and $\ell_L\approx L$.
That is, for most parameter combinations $(M,N,\alpha)$ we expect $1< \ell_L < L$.

Below we look for the scaling properties of the 
eigenfunctions of ${\bf A}$ through the {\it scaled localization length}
\begin{equation}
\beta = \frac{\ell_L}{L} ,
\label{betaS}
\end{equation}
which can take values in the range $(0,1]$.
Indeed, outstandingly, it has been found that the eigenfunction properties of
diluted banded random matrices, both
Hermitian~\cite{MFMR17b} and non-Hermitian~\cite{HTM24}, 
as well as for non-directed multilayer networks~\cite{MFMR17a},
obey the scaling function
\begin{equation}
\beta = \frac{x^*}{1+x^*} ,
\label{betax*}
\end{equation}
where
\begin{equation}
x^* \equiv \gamma x^\delta ,
\label{x*}
\end{equation}
$\gamma\equiv\gamma(\alpha)$, $\delta\equiv\delta(\alpha)\sim 1$,
and
\begin{equation}
x = b_{\mbox{\tiny eff}}^2/L .
\label{x}
\end{equation}
Above, $b_{\mbox{\tiny eff}}$ is the effective bandwidth of the corresponding
diluted banded random matrix of size $L$.

It is relevant to add that the scaling~(\ref{betax*}) with $\gamma\sim 1$ and $\delta=1$ also works 
for non-diluted banded random matrices, see Refs.~\cite{CMI90,EE90,FM91,I95,FM92,MF93,FM93,FM94}, 
as well as for the kicked-rotator model~\cite{CGIS90,I90,I95} (a quantum-chaotic system characterized by 
a random-like banded Hamiltonian matrix), the one-dimensional Anderson model, and the Lloyd 
model~\cite{CGIFM92}.

\begin{figure*}[ht]
\centering
\includegraphics[width=0.65\textwidth]{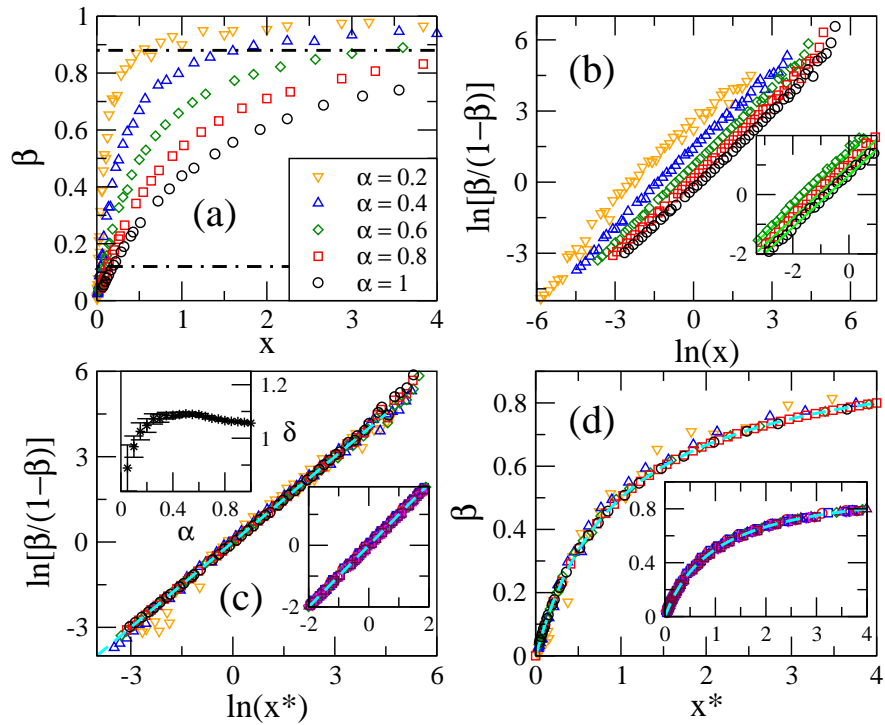}
\caption{
(a) Scaled localization length of eigenfunctions $\beta$ as a function of $x=b_{\mbox{\tiny eff}}^2/L$, 
$b_{\mbox{\tiny eff}}=2N\alpha$, for multilayer directed random networks 
characterized by the sparsity $\alpha$. Several combinations of $(M,N)$ are used for each value of $\alpha$.
Horizontal dot-dashed lines at $\beta\approx 0.12$ and 0.88 are shown as a reference, see the text.
(b) Logarithm of $\beta/(1-\beta)$ as a function of $\ln(x)$. 
Inset: Enlargement in the range $\ln[\beta/(1-\beta)]=[-2,2]$ including data for 
$\alpha=0.6$, 0.8, and 1. Green-dashed lines are fittings of the data with Eq.~(\ref{betascaling2}).
(c) Logarithm of $\beta/(1-\beta)$ as a function of $\ln(x^*)$ [see Eq.~(\ref{x*})]. 
Upper inset: Power $\delta$, obtained from fittings of the curves $\ln[\beta/(1-\beta)]$ vs.~$\ln(x)$ 
in the range $\ln[\beta/(1-\beta)]=[-2,2]$ with Eq.~(\ref{betascaling2}), as a function of $\alpha$. 
Lower Inset: Enlargement in the range $\ln[\beta/(1-\beta)]=[-2,2]$ including curves for 
$\alpha\in [0.5,1]$ in steps of $0.05$. Cyan-dashed lines in main panel and lower inset 
are Eq.~(\ref{betascaling3}).
(d) $\beta$ as a function of $x^*$. 
Inset: Data for $\alpha\in [0.5,1]$ in steps of $0.05$. Cyan-dashed lines in main 
panel and inset are Eq.~(\ref{betax*}).
}
\label{Fig01}
\end{figure*}
\begin{figure*}[ht]
\centering
\includegraphics[width=0.6\textwidth]{Fig02.eps}
\caption{
(a) $\left\langle \overline{\mbox{IPR}} \right\rangle$, 
(b) $\left\langle \overline{r}_\mathbb{C} \right\rangle$ and
(c) $\left\langle \overline{r}_\mathbb{SV} \right\rangle$
as a function of $x$ for multilayer directed random networks 
characterized by the sparsity $\alpha$. 
(d) $\left\langle \overline{\mbox{IPR}} \right\rangle$, 
(e) $\left\langle \overline{r}_\mathbb{C} \right\rangle$ and
(f) $\left\langle \overline{r}_\mathbb{SV} \right\rangle$
as a function of $x^*$.
Insets in (d-f): 
Data for $\alpha\in [0.5,1]$ in steps of $0.05$. Dashed lines are Eq.~(\ref{betax*}).
}
\label{Fig02}
\end{figure*}
\begin{figure*}[ht]
\centering
\includegraphics[width=0.65\textwidth]{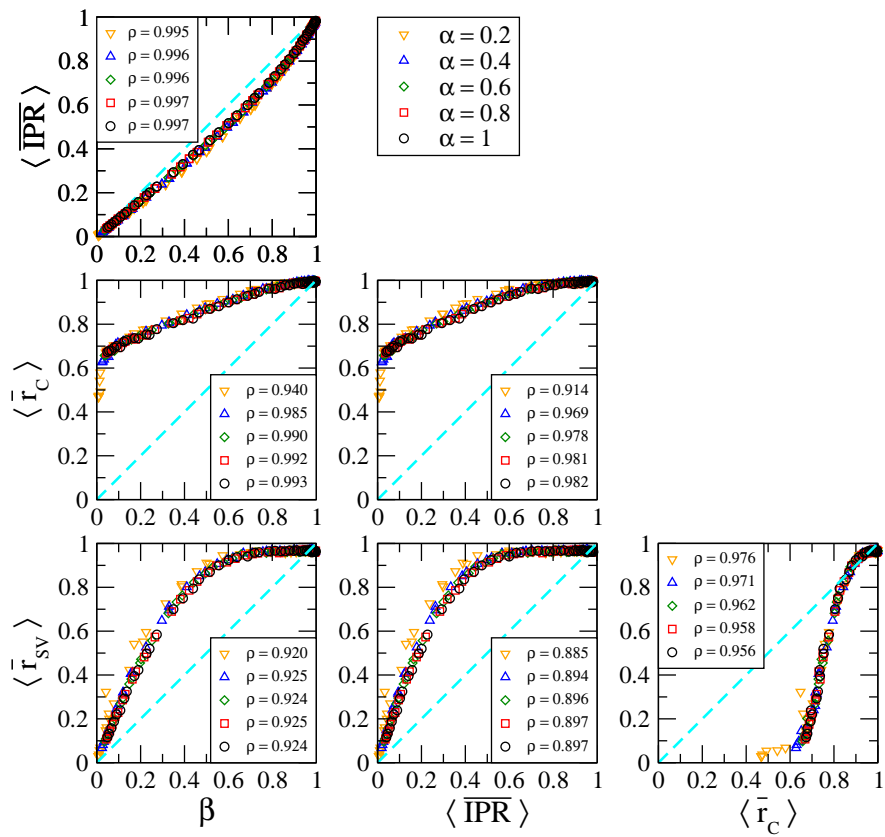}
\caption{
Scatter plots between $\beta$, $\left\langle \overline{\mbox{IPR}} \right\rangle$, 
$\left\langle \overline{r}_\mathbb{C} \right\rangle$ and
$\left\langle \overline{r}_\mathbb{SV} \right\rangle$ for multilayer directed random networks. 
The Pearson's correlation coefficients $\rho$ are reported in the corresponding panels.
Cyan-dashed lines, plotted to guide the eye, are the identity. 
}
\label{Fig03}
\end{figure*}

\subsection{Scaling analysis of the localization length of eigenfunctions}

In Fig.~\ref{Fig01}(a) we present the scaled localization length $\beta$ as a function of 
$x=b_{\mbox{\tiny eff}}^2/L$, with
\begin{equation}
b_{\mbox{\tiny eff}} = 2N\alpha ,
\label{beff}
\end{equation}
for ensembles of multilayer directed random networks characterized by the sparsity $\alpha$. 
We stress that $b_{\mbox{\tiny eff}}$ is the average number of non-vanishing elements per 
adjacency-matrix row.
In Fig.~\ref{Fig01}(a) for every value of $\alpha$ we consider several combinations of $M$ 
and $N$ to cover a wide range of values of $x$.

We observe that the curves of $\beta$ vs.~$x$ show similar functional forms however 
clearly affected by the sparsity $\alpha$: For a fixed $x$, the smaller the value of $\alpha$ 
the larger the value of $\beta$ is. 
This panorama is equivalent to that reported for diluted banded random matrices, both
Hermitian~\cite{MFMR17b} and non-Hermitian~\cite{HTM24}, 
as well as for non-directed multilayer networks~\cite{MFMR17a}.
In addition, in Fig.~\ref{Fig01}(b) the logarithm of $\beta/(1-\beta)$ as a function of $\ln(x)$ is also 
presented. The quantity $\beta/(1-\beta)$ was useful in the study of the scaling properties of  
diluted banded random matrices and of non-directed multilayer networks because 
$\beta/(1-\beta) \propto x^\delta$ implies that a 
plot of $\ln[\beta/(1-\beta)]$  vs.~$\ln(x)$ is a straight line with slope $\delta$. 
In fact, from Fig.~\ref{Fig01}(b) we confirm that plots of 
$\ln[\beta/(1-\beta)]$ vs.~$\ln(x)$ are straight lines (in a wide range of $x$) with 
a slope that depends (slightly but detectably) on the sparsity $\alpha$. 
Consequently, we put to test the scaling law 
\begin{equation}
\frac{\beta}{1-\beta} = \gamma x^\delta ,
\label{betascaling2}
\end{equation}
where $\gamma\equiv\gamma(\alpha)$ and $\delta\equiv\delta(\alpha)$. 
Notice that scaling~(\ref{betascaling2}) is equivalent to scaling~(\ref{betax*}).

Indeed, Eq.~(\ref{betascaling2})
describes well the data, mainly in the range $\ln[\beta/(1-\beta)]=[-2,2]$, as can be 
seen in the inset of Fig.~\ref{Fig01}(b) where we show the numerical data for 
$\alpha=0.6$, 0.8 and 1 including fittings with Eq.~(\ref{betascaling2}).
We stress that the range $\ln[\beta/(1-\beta)]=[-2,2]$ corresponds to a
reasonable wide range of $\beta$ values, $\beta\approx[0.12,0.88]$, whose bounds
are indicated with horizontal dot-dashed lines in Fig.~\ref{Fig01}(a).
Finally, we notice that the power $\delta$, obtained from the fittings of the data 
using Eq.~(\ref{betascaling2}), is very close to unity for all the sparsity
values we consider here (see the upper inset of Fig.~\ref{Fig01}(c)).

Therefore, from the analysis of the data in Figs.~\ref{Fig01}(a,b), we are able to write down 
a {\it universal scaling function} for the scaled localization length $\beta$ of the
eigenfunctions of multilayer directed random network model as
\begin{equation}
\beta/(1-\beta) = x^* ,
\label{betascaling3}
\end{equation}
where the scaling parameter $x^*$, see Eq.~(\ref{x*}), as a function of the multilayer network 
parameters, is given by
\begin{equation}
x^* \equiv \gamma \left( \frac{4N\alpha^2}{M} \right)^\delta .
\label{x*multilayer}
\end{equation}
To validate Eq.~(\ref{betascaling3}) in Fig.~\ref{Fig01}(c) we present again the
data for $\ln[\beta/(1-\beta)]$ shown in Fig.~\ref{Fig01}(b) but now as a function 
of $\ln(x^*)$. We do observe that curves for different values of $\alpha$ fall on 
top of Eq.~(\ref{betascaling3}) for a wide range of the variable $x^*$.
Moreover, the collapse of the numerical data is excellent in the range 
$\ln[\beta/(1-\beta)]=[-2,2]$ for $\alpha\ge 0.5$, as
shown in the lower inset of Fig.~\ref{Fig01}(c).

Finally, in Fig.~\ref{Fig01}(d) we confirm the validity of Eq.~(\ref{betax*}).
We emphasize that the universal scaling given in Eq.~(\ref{betax*})
extends outside the range $\beta\approx[0.12,0.88]$, for which Eq.~(\ref{betascaling2})
was shown to be valid, see the main panel of Fig.~\ref{Fig01}(c). Furthermore, 
the collapse of the numerical data on top of Eq.~(\ref{betax*}) is remarkably good for 
$\alpha\ge 0.5$, as shown in the inset of Fig.~\ref{Fig01}(d).

\begin{figure*}[ht]
\centering
\includegraphics[width=0.65\textwidth]{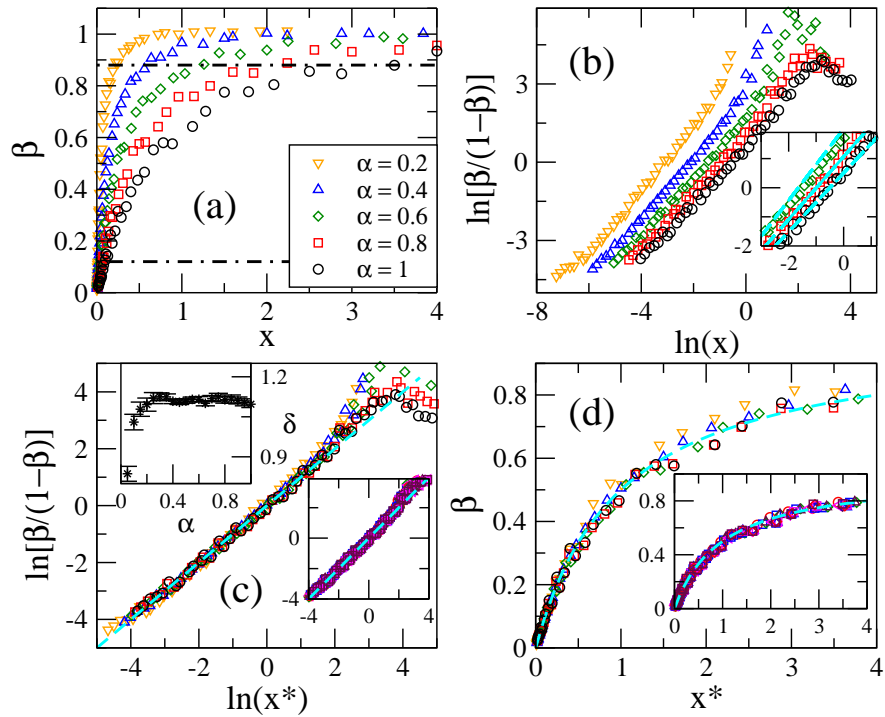}
\caption{
(a) Scaled localization length of eigenfunctions $\beta$ as a function of $x=b_{\mbox{\tiny eff}}^2/L$, 
$b_{\mbox{\tiny eff}}=N\alpha$, for multiplex directed random networks 
characterized by the sparsity $\alpha$. Several combinations of $(M,N)$ are used for each value of $\alpha$.
Horizontal dot-dashed lines at $\beta\approx 0.12$ and 0.88 are shown as a reference, see the text.
(b) Logarithm of $\beta/(1-\beta)$ as a function of $\ln(x)$. 
Inset: Enlargement in the range $\ln[\beta/(1-\beta)]=[-2,2]$ including data for 
$\alpha=0.6$, 0.8, and 1. Green-dashed lines are fittings of the data with Eq.~(\ref{betascaling2}).
(c) Logarithm of $\beta/(1-\beta)$ as a function of $\ln(x^*)$ [see Eq.~(\ref{x*multiplex})]. 
Upper inset: Power $\delta$, obtained from the fittings of the curves $\ln[\beta/(1-\beta)]$ vs.~$\ln(x)$ 
in the range $\ln[\beta/(1-\beta)]=[-2,2]$ with Eq.~(\ref{betascaling2}), as a function of $\alpha$. 
Lower Inset: Enlargement in the range $\ln[\beta/(1-\beta)]=[-2,2]$ including curves for 
$\alpha\in [0.5,1]$ in steps of $0.05$. Cyan-dashed lines in main panel and lower inset 
are Eq.~(\ref{betascaling3}).
(d) $\beta$ as a function of $x^*$. 
Inset: Data for $\alpha\in [0.5,1]$ in steps of $0.05$. Cyan-dashed lines in main 
panel and inset are Eq.~(\ref{betax*}).
}
\label{Fig04}
\end{figure*}
\begin{figure*}[ht]
\centering
\includegraphics[width=0.6\textwidth]{Fig05.eps}
\caption{
(a) $\left\langle \overline{\mbox{IPR}} \right\rangle$, 
(b) $\left\langle \overline{r}_\mathbb{C} \right\rangle$ and
(c) $\left\langle \overline{r}_\mathbb{SV} \right\rangle$
as a function of $x$ for multiplex directed random networks 
characterized by the sparsity $\alpha$. 
(d) $\left\langle \overline{\mbox{IPR}} \right\rangle$, 
(e) $\left\langle \overline{r}_\mathbb{C} \right\rangle$ and
(f) $\left\langle \overline{r}_\mathbb{SV} \right\rangle$
as a function of $x^*$.
Insets in (d-f): 
Data for $\alpha\in [0.5,1]$ in steps of $0.05$. Dashed lines are Eq.~(\ref{betax*}).
}
\label{Fig05}
\end{figure*}
\begin{figure*}[ht]
\centering
\includegraphics[width=0.65\textwidth]{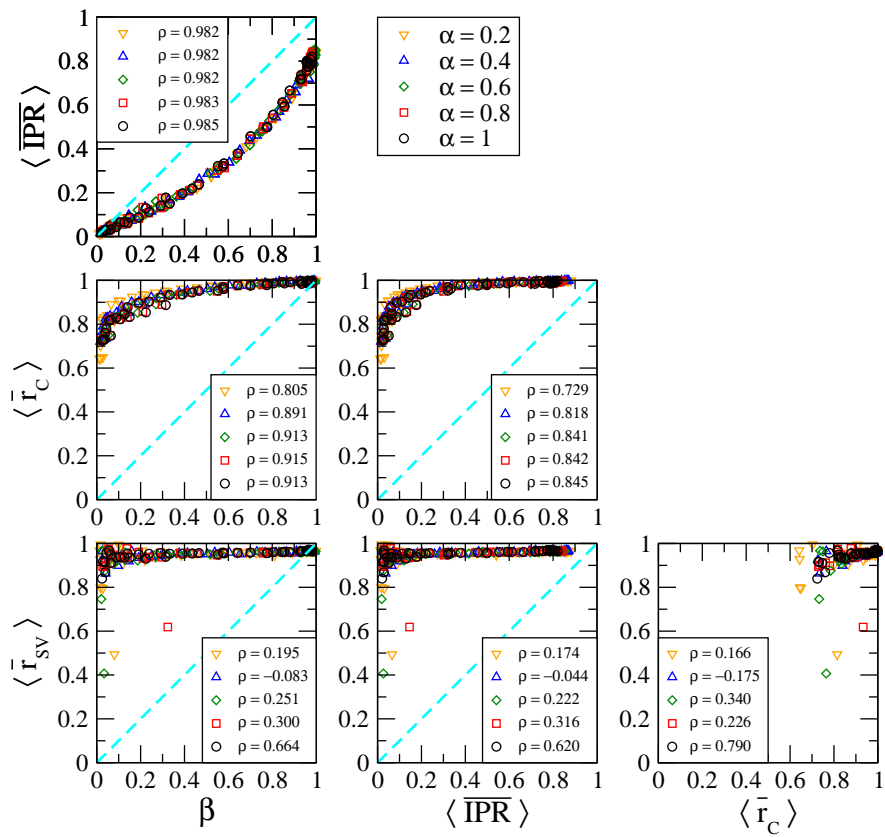}
\caption{
Scatter plots between $\beta$, $\left\langle \overline{\mbox{IPR}} \right\rangle$, 
$\left\langle \overline{r}_\mathbb{C} \right\rangle$ and
$\left\langle \overline{r}_\mathbb{SV} \right\rangle$ for multiplex directed random networks. 
The Pearson's correlation coefficients $\rho$ are reported in the corresponding panels.
Cyan-dashed lines, plotted to guide the eye, are the identity. 
}
\label{Fig06}
\end{figure*}

\subsection{Scaling of additional RMT measures}

In what follows we complete the analysis of eigenfunction and spectral properties of multilayer 
directed random networks by computing 
the average inverse participation ratios $\left\langle \mbox{IPR} \right\rangle$ as well
as the average ratios $\left\langle r_\mathbb{C} \right\rangle$ and 
$\left\langle r_{\tbox{SV}} \right\rangle$, see Eqs.~(\ref{IPR}-\ref{rSV}).
Moreover, we conveniently normalize these averages as follows:
\begin{equation}
\langle \overline{\mathrm{IPR}}\rangle= \frac{\langle \mathrm{IPR} \rangle-\mathrm{IPR}_{\mathrm{PE}}}{\mathrm{IPR}_{\mathrm{RGE}}-\mathrm{IPR}_{\mathrm{PE}}} ,
\label{anIPR}
\end{equation}
\begin{equation}
\langle \overline{r}_\mathbb{C}\rangle= \frac{\langle r_\mathbb{C}  \rangle-r_{\mathbb{C}_{\mathrm{PE}}}}{r_{\mathbb{C}_{\mathrm{RGE}}}-r_{\mathbb{C}_{\mathrm{PE}}}} 
\label{anrC}
\end{equation}
and
\begin{equation}
\langle \overline{r}_{\tbox{SV}}\rangle= \frac{\langle r_{\tbox{SV}}  \rangle-r_{{\tbox{SV}}_{\mathrm{PE}}}}{r_{{\tbox{SV}}_{\mathrm{RGE}}}-r_{{\tbox{SV}}_{\mathrm{PE}}}} ,
\label{anrSV}
\end{equation}
such that they all take values in the interval $[0,1]$, so they can be directly compared with $\beta$.
The reference values used in Eqs.~(\ref{anIPR}-\ref{anrSV}), 
corresponding to the PE and the RGE, are reported in Table~\ref{T1}.

\begin{table}[b!]
\caption{
Reference values of the RMT measures for the Poisson ensemble 
and the real Ginibre ensemble used in Eqs.~(\ref{anIPR}-\ref{anrSV}).
}
\label{T1}
\begin{tabular}{ c | c | c | c  }  
\hline
 & $\mathrm{IPR}$  & $r_\mathbb{C}$ & $r_{\tbox{SV}}$ \\
\hline
PE    & 1 & 0.5~\cite{PRRCM20} & 0.386~\cite{ABGR13} \\
RGE & $N$/2.04~\cite{PM23} & 0.737~\cite{PRRCM20} & 0.536~\cite{ABGR13} \\
\hline
\end{tabular}
\end{table}

Then, in Figs.~\ref{Fig02}(a-c) we plot the normalized measures 
$\left\langle \overline{\mbox{IPR}} \right\rangle$, 
$\left\langle \overline{r}_\mathbb{C} \right\rangle$ and
$\left\langle \overline{r}_{\tbox{SV}} \right\rangle$, respectively,
as a function of $x$ for multilayer 
directed random networks characterized by the sparsity $\alpha$.
The panorama shown in Figs.~\ref{Fig02}(a-c) for $\left\langle \overline{\mbox{IPR}} \right\rangle$, 
$\left\langle \overline{r}_\mathbb{C} \right\rangle$ and
$\left\langle \overline{r}_{\tbox{SV}} \right\rangle$ is equivalent to that observed for
$\beta$ in Fig.~\ref{Fig01}(a): 
The curves of $\left\langle \overline{X} \right\rangle$ vs.~$x$ show similar functional forms however 
clearly affected by the sparsity $\alpha$. 
Here, $X$ represents $\mbox{IPR}$, $r_\mathbb{C}$ and $r_{\tbox{SV}}$.
Also, for a fixed $x$, the smaller the value of $\alpha$ the larger the value of $\left\langle \overline{X} \right\rangle$. 
This observations allows us to surmise that the scaling parameter of $\beta$, $x^*$, may also
serve as scaling parameter of $\left\langle \overline{X} \right\rangle$.
To verify this assumption, in Figs.~\ref{Fig02}(d-f) we plot again  
$\left\langle \overline{\mbox{IPR}} \right\rangle$, 
$\left\langle \overline{r}_\mathbb{C} \right\rangle$ and
$\left\langle \overline{r}_{\tbox{SV}} \right\rangle$, respectively,
but now as a function of $x^*$.
Indeed, since the curves $\left\langle \overline{X} \right\rangle$ vs.~$x^*$ fall one on top of the other 
mainly for $\alpha\ge 0.5$, see the corresponding insets, we conclude that $x^*$ scales 
$\left\langle \overline{\mbox{IPR}} \right\rangle$, 
$\left\langle \overline{r}_\mathbb{C} \right\rangle$ and
$\left\langle \overline{r}_{\tbox{SV}} \right\rangle$
as good as it scales $\beta$.
From Fig.~\ref{Fig02} we also observe that the curves
$\left\langle \overline{r}_\mathbb{C} \right\rangle$ vs.~$x^*$ and 
$\left\langle \overline{r}_{\tbox{SV}} \right\rangle$ vs.~$x^*$
are above Eq.~(\ref{betax*}), which is included as dashed lines.
This also means that the spectral properties of multilayer 
directed random networks approach the RGE limit faster 
than the eigenfunction properties.
Also, while $\left\langle \overline{\mbox{IPR}} \right\rangle$ and
$\left\langle \overline{r}_{\tbox{SV}} \right\rangle$ characterizes the complete transition from 
localization (i.e.~$\left\langle \overline{\mbox{IPR}} \right\rangle\approx 0$ and
$\left\langle \overline{r}_{\tbox{SV}} \right\rangle\approx 0$ when $x*\to 0$) to 
delocalization (i.e.~$\left\langle \overline{\mbox{IPR}} \right\rangle\to 1$ and
$\left\langle \overline{r}_{\tbox{SV}} \right\rangle\to 1$ when $x*\gg 1$), 
$\left\langle \overline{r}_\mathbb{C} \right\rangle$ does not span the full range 
$0<\left\langle \overline{r}_\mathbb{C} \right\rangle<1$, 
see Figs.~\ref{Fig02}(e). So, $\left\langle \overline{r}_\mathbb{C} \right\rangle$ is a
poor measure to characterize multilayer .directed random networks

Moreover, remarkably, $\left\langle \overline{\mbox{IPR}} \right\rangle$ closely follows the same
scaling law as $\beta$, so we can write
\begin{equation}
\left\langle \overline{\mbox{IPR}} \right\rangle \approx \frac{x^*}{1+x^*} ,
\label{IPRx*}
\end{equation}
see the inset in Fig.~\ref{Fig02}(d). 
In fact, this also happens for Hermitian diluted banded random matrices~\cite{HTM24}.

From Figs.~\ref{Fig01} and~\ref{Fig02} we can also see that all quantities ($\beta$, 
$\left\langle \overline{\mbox{IPR}} \right\rangle$, 
$\left\langle \overline{r}_\mathbb{C} \right\rangle$ and
$\left\langle \overline{r}_{\tbox{SV}} \right\rangle$)
appear to be highly correlated, mainly $\beta$ and
$\left\langle \overline{\mbox{IPR}} \right\rangle$. Therefore, in Fig.~\ref{Fig03} we present
scatter plots between $\beta$, $\left\langle \overline{\mbox{IPR}} \right\rangle$, 
$\left\langle \overline{r}_\mathbb{C} \right\rangle$ and
$\left\langle \overline{r}_{\tbox{SV}} \right\rangle$ for multilayer 
directed random networks for several values of $\alpha$, where the high correlation between them is evident.
To quantify the correlation among these quantities, in the panels of Fig.~\ref{Fig03} we report the
Pearson's correlation coefficient $\rho$ between all the RMT measures we computed.
Specifically, we contrast data sets characterized by the same values of $\alpha$. Note that in all cases
we obtain $\rho>0.9$, meaning a relatively large correlation among all measures.
We stress that since $\beta$ and $\left\langle \overline{\mbox{IPR}} \right\rangle$ obey approximately the 
same scaling law, the correlation among them is specially strong, see the upper-left panel in Fig.~\ref{Fig03}.

\section{Scaling analysis of multiplex directed random networks}

We now turn our attention to multiplex directed random networks whose adjacency matrix 
is given in Eq.~(\ref{eq:A_line}). We follow the same methodology as in the multilayer 
case presented in the previous Section. 

Thus, in Fig.~\ref{Fig04}(a) we first present curves of $\beta$ vs.~$x$ as given in 
Eqs.~(\ref{betaS}) and (\ref{x}), respectively; however, we now define
$b_{\mbox{\tiny eff}}$ as 
\begin{equation}
b_{\mbox{\tiny eff}} = N\alpha ,
\label{beff2}
\end{equation}
which is the average number of non-vanishing elements per row inside the 
adjacency-matrix band in the multiplex setup. 
From Fig.~\ref{Fig04}(a) we observe that the curves of $\beta$ vs.~$x$ have
functional forms similar to those for the multilayer model (compare with Fig.~\ref{Fig01}(a));
however, with larger values of $\beta$ for given values of $x$. 
Moreover, in Fig.~\ref{Fig04}(b) we show the logarithm of $\beta/(1-\beta)$ as a function of $\ln(x)$. 
As in the multilayer case, here we observe that 
plots of $\ln[\beta/(1-\beta)]$ vs.~$\ln(x)$ are straight lines mainly in the range
$\ln[\beta/(1-\beta)]=[-2,2]$ with a slope that depends on the sparsity $\alpha$.
We indicate the bounds of this range with horizontal dot-dashed lines in Fig.~\ref{Fig04}(a).
Therefore, the scaling law of Eq.~(\ref{betascaling2}) should also be valid here. Indeed, in the upper inset of 
Fig.~\ref{Fig04}(c) we report the power $\delta$ obtained from fittings of the data with 
Eq.~(\ref{betascaling2}).

In order to validate the scaling hypothesis of Eq.~(\ref{betascaling2}) for the node-aligned multiplex 
directed setup, in Fig.~\ref{Fig04}(c) we present the data for $\ln[\beta/(1-\beta)]$ shown in 
Fig.~\ref{Fig04}(b), but now as a function of $\ln(x^*)$. We observe that curves for different 
values of  $\alpha$ fall on top of Eq.~(\ref{betascaling3}) for a wide range of the variable $x^*$,
which for multiplex directed random networks is given by
\begin{equation}
x^* \equiv \gamma \left( \frac{N\alpha^2}{M} \right)^\delta .
\label{x*multiplex}
\end{equation}
Moreover, the collapse of the numerical data on top of 
Eq.~(\ref{betascaling3}) is excellent in the range $\ln[\beta/(1-\beta)]=[-2,2]$ for $\alpha\ge 0.5$, as 
shown in the lower inset of Fig.~\ref{Fig04}(c).
Finally, in Fig.~\ref{Fig04}(d) we confirm the validity of Eq.~(\ref{betax*}) which is 
as good here as for the multilayer case. We emphasize that 
the collapse of the numerical data on top of Eq.~(\ref{betax*}) is remarkably good for 
$\alpha\ge 0.5$, as shown in the inset of Fig.~\ref{Fig04}(d).

Then, in Fig.~\ref{Fig05} we report additional RMT measures to characterize eigenfunction and spectral 
properties of multiplex directed random networks:
(a) $\left\langle \overline{\mbox{IPR}} \right\rangle$, 
(b) $\left\langle \overline{r}_\mathbb{C} \right\rangle$ and
(c) $\left\langle \overline{r}_{\tbox{SV}} \right\rangle$.
Note that Fig.~\ref{Fig05} is equivalent to Fig.~\ref{Fig02} for multilayer directed random networks.

From Fig.~\ref{Fig05} we observe that $x^*$ works reasonably well as scaling parameter of 
$\left\langle \overline{\mbox{IPR}} \right\rangle$, 
$\left\langle \overline{r}_\mathbb{C} \right\rangle$ and
$\left\langle  \overline{r}_{\tbox{SV}} \right\rangle$ of multiplex directed random networks, see Figs.~\ref{Fig05}(d-f).
Particularly for $\alpha\ge 0.5$, where all curves $\left\langle \overline{X} \right\rangle$ vs.~$x^*$ fall 
one on top of the other; see the corresponding insets.
Here, $X$ represents $\mbox{IPR}$, $r_\mathbb{C}$ and $r_{\tbox{SV}}$.
Also, in contrast with the multilayer case, here $\left\langle \overline{r}_\mathbb{C} \right\rangle$
appears to be a better measure than $\left\langle  \overline{r}_{\tbox{SV}} \right\rangle$ to
characterize the delocalization transition of multiplex directed random networks.
That is, $\left\langle  \overline{r}_{\tbox{SV}} \right\rangle$ is relatively close to 1 even for
$x^*\to 0$. This is a consequence of the complete graph structure of the multiplex which
prevails even for $x^*\to 0$.

Finally, in Fig.~\ref{Fig06} we present the scatter plots between $\beta$, 
$\left\langle \overline{\mbox{IPR}} \right\rangle$, 
$\left\langle \overline{r}_\mathbb{C} \right\rangle$ and
$\left\langle  \overline{r}_{\tbox{SV}} \right\rangle$ for multiplex directed random networks. 
We also include the Pearson's correlation coefficients $\rho$ in the corresponding panels.
Here, in contrast with multilayer directed random networks, correlations are large among 
$\beta$, $\left\langle \overline{\mbox{IPR}} \right\rangle$ and
$\left\langle \overline{r}_\mathbb{C} \right\rangle$ only.

\section{Conclusions}

In this work, by means of the scaling analysis within a random matrix theory (RMT) approach, 
we have demonstrated that the normalized localization 
length $\beta$ of the eigenfunctions of multilayer directed random networks (composed by $M$ layers 
of size $N$) scales as $x^*/(1+x^*)$; see Figs.~\ref{Fig01}(d) and~\ref{Fig04}(d).
Here $x^*=\gamma(b_{\mbox{\tiny eff}}^2/L)^\delta$, $\delta\sim 1$ and $b_{\mbox{\tiny eff}}$ 
being the effective bandwidth of the adjacency matrix of the network of size $L=M\times N$.
The quantities $\gamma$, $\delta$, and $b_{\mbox{\tiny eff}}$ depend on both intra-- and inter--layer
connectivity, here quantified by the parameter $\alpha$.
We also showed that other eigenfunction and spectral RMT measures scale with $x^*$; see 
Figs.~\ref{Fig02}(d-f) and~\ref{Fig05}(d-f).
It is relevant to add that our results are robust for multilayer as well as multiplex setups, see e.g.~Fig.~\ref{Fig00}.

We believe that the scaling given by Eq.~(\ref{betax*}), which turns out to also apply to non-directed
random networks~\cite{MFMR17a}, may be used to either predict or tune the transport properties of multilayer 
random networks, since eigenfunction localization properties have a direct effect on transport.
That is, localized eigenfunctions (i.e.~$\beta\sim 0$ or $x^*\ll 1$) are characteristic of an insulating regime 
while delocalized (or extended)
eigenfunctions (i.e.~$\beta\sim 1$ or $x^*\gg 1$) are proper of a metallic (diffusive) regime.
In this respect $x^*=1$, i.e.~$\beta\sim 1/2$, represents a natural localization--to--delocalization crossover point.
More specifically, given a multilayer network characterized by a given value of $x^*$, its eigenfunctions could 
be made more/less localized by decreasing/increasing $x^*$.
For this task, one has several possibilities (depending on the nature of the network), see 
Eqs.~(\ref{x*multilayer}) and~(\ref{x*multiplex}):
decrease/increase the layer size $N$, decrease/increase the sparsity $\alpha$, or add/remove layers.

We hope our results motivate further numerical and theoretical studies.

\section*{Acknowledgements}

J.A.M.-B. thanks support from VIEP-BUAP (Grant No.~100405811-VIEP2024), Mexico.

\appendix
\renewcommand{\thefigure}{A\arabic{figure}}
\setcounter{figure}{0}

\section{Multiplex directed random networks. The case of $\omega<1$}
\label{app}

\begin{figure*}[ht]
\centering
\includegraphics[width=0.65\textwidth]{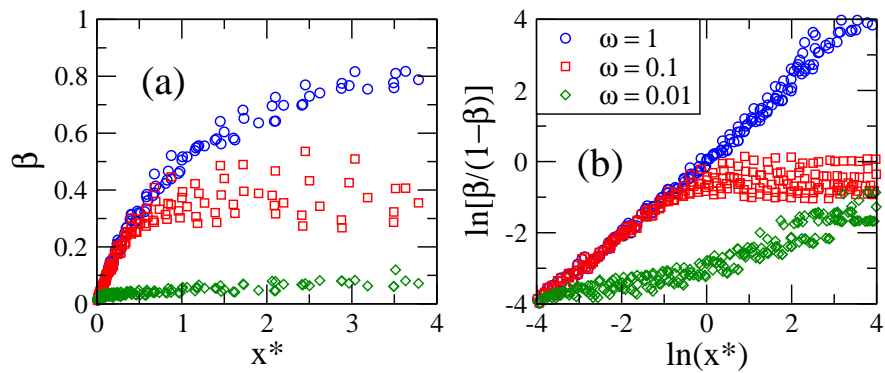}
\caption{
(a) Scaled localization length of eigenfunctions $\beta$ as a function of $x^*$, see Eq.~(\ref{x*multiplex}),
for multiplex directed random networks characterized by the interlayer coupling strength $\omega$. 
Several combinations of $(M,N,\alpha)$ are used.
(b) Logarithm of $\beta/(1-\beta)$ as a function of $\ln(x^*)$. 
}
\label{Fig07}
\end{figure*}

Here, we explore the case of multiplex directed random networks with a strength of interlayer
edges, $\omega$, different from unity; that is, now the corresponding adjacency matrix has the form
\begin{equation}
{\bf A} =
\left(
\begin{array}{ccccc}
A^{(1)} & \omega I & \omega I & \cdots & \omega I \\
\omega I & A^{(2)} & \omega I & \ & \omega I \\
\omega I & \omega I & A^{(3)} & \ & \omega I \\
\vdots & \ & \ & \ddots & \omega I \\
\omega I & \omega I & \omega I & \omega I & A^{(M)}
\end{array}
\right) \ .
\label{eq:A_muxw}
\end{equation}
Here, as well as in Ref.~\cite{RJ23} we consider $\omega<1$.

In Fig.~\ref{Fig07}(a) we present the scaled localization length of eigenfunctions $\beta$ as a function of 
$x^*$, see Eq.~(\ref{x*multiplex}), for multiplex directed random networks characterized by the interlayer 
coupling strength $\omega=0.1$ (red symbols) and $\omega=0.01$ (green symbols).
For comparison purposes we also include the case of $\omega=1$ (blue symbols).
In Fig.~\ref{Fig07}(b) we show the same data but as $\ln[\beta/(1-\beta)]$ vs.~$\ln(x^*)$.

Form Fig.~\ref{Fig07} we observe that
(i) For fixed $x^*$, $\beta$ decreases for decreasing $\omega$. This could have been anticipated since 
for $\omega\ll 1$ the layers become uncoupled and localization inside the layers is expected. This is also 
in agreement with Ref.~\cite{RJ23}.
(ii) For $\omega<1$, $\beta$ does not scale with $x^*$. This means that a proper scaling analysis should
be done in order to define the correct scaling parameter that should also include $\omega$.
However, this is not the purpose of the present work, so we do not perform the corresponding scaling 
analysis here.

\bibliographystyle{plainnat}

\end{document}